\documentclass[11pt]{article}
 
\usepackage{epsfig}
\begin{document}

\baselineskip=14pt plus 0.2pt minus 0.2pt
\lineskip=14pt plus 0.2pt minus 0.2pt

\newcommand{\be}{\begin{equation}}
\newcommand{\ee}{\end{equation}}
\newcommand{\bea}{\begin{eqnarray}}
\newcommand{\eea}{\end{eqnarray}}
\newcommand{\da}{\dagger}
\newcommand{\dg}[1]{\mbox{${#1}^{\dagger}$}}
\newcommand{\hlf}{\mbox{$1\over2$}}
\newcommand{\lfrac}[2]{\mbox{${#1}\over{#2}$}}
\newcommand{\scsz}[1]{\mbox{\scriptsize ${#1}$}}
\newcommand{\tsz}[1]{\mbox{\tiny ${#1}$}}

\begin{flushright}
hep-th/0005281 \\
LA-UR-00-2368 \\
\end{flushright} 

\begin{center}
\large{\bf A Simple Volcano Potential with an Analytic, Zero-Energy, 
Ground State}
 
\vspace{0.25in}

\normalsize
\bigskip

Michael Martin Nieto\footnote{\noindent  Email:  
mmn@lanl.gov}\\
{\it Theoretical Division (MS-B285), Los Alamos National Laboratory\\
University of California\\
Los Alamos, New Mexico 87545, U.S.A. \\}
 
\normalsize

\vskip 20pt
\today

\vspace{0.3in}

{ABSTRACT}

\end{center}

\begin{quotation}

We describe a simple volcano potential, which is supersymmetric and has
an analytic, zero-energy, ground state.  (The KK modes are also analytic.)  
It is an interior harmonic
oscillator potential properly matched to an exterior angular momentum-like
tail.  Special cases are given to elucidate the physics, which may be
intuitively useful in studies of higher-dimensional gravity.

\vspace{0.25in}

\end{quotation}

\newpage

\baselineskip=.33in

\section{Introduction}

There has been recent excitement about the observation that extra dimensions
in gravity could be of macroscopic size and yet not be in conflict with
experiment \cite{arkani}.  This led to the proposal \cite{randall} 
of  a scenario where one can have extra dimensions that are not 
compactified at all, but yet have negligible effect on normal-scale 
gravitational physics \cite{randall}.  In one extra dimension, 
the problem effectively reduces to a 
Schr\"odinger-like equation for the graviton mass squared, $m^2$.  
If such a model can yield a zero-energy ground state
which is bound, then the massless graviton can be confined to a local 
region even though its quantum equation in the higher dimension is valid 
over the whole line:     
\be
\left[-\frac{d^2}{dx^2} + V(x)\right]\psi_m(x) =m^2, 
\ee
One of the prime candidates for such a system is a  
``volcano" potential \cite{randall,csaki}.  

Such a  model should also be supersymmetric. 
Recall that to be supersymmetric \cite{mmns}, the zero-energy 
ground state and the potential are  given by   
\bea
\psi_0(x)&=&N \exp[-W(x)],  \label{psiasw} \\
V(x) &=& \left[ W'(x)\right]^2 - W''(x),  \label {vasw}   
\eea
where $N$ is the normalization constant and $W$ is the superpotential.   

A fair amount of effort has gone into  studies of 
potentials with the above properties \cite{randall,csaki,gremm,dewolfe}. 
These `volcano-like'' potentials  include  
i) a polynomial tail with a negative delta function at the origin
\cite{randall}; ii) one from a superpotential $W=\ln[k^2x^2+1]$
\cite{csaki}; and iii) one composed of a 
square-well box at the origin matched to two square steps on the sides
\cite{csaki}.   Of course, none of these is a perfect toy model.  The best
understood analytically are the delta function and box potentials. 
(The latter, however, does not come from a superpotential).  

Here, we will construct a supersymmetric potential which is simple 
and has a simple, exactly analytic solution for the zero-energy ground
state. The excited states are complicated, but are known special 
functions of mathematical physics.  Perhaps such a potential can  
be useful in studying the physics of unconfined extra dimensions.  


\section{The potential}   

The idea is to match, in the proper manner,  
a simple, exactly solvable, central, crater potential to a simple, 
exactly solvable, tail potential. 
The ``appropriate match''and ``simple'' will yield a particular step
function jump from the crater to the tail and a particular matching point.  
Motivated by physical intuition and work on $E=0$ bound state
solutions for singular potentials   \cite{daboulq,daboulc}, we take the
exterior potential to be an ``angular-momentum-like'' tail. This is 
the tail with the simplest analytic behavior.   We take the central
potential to be an harmonic oscillator.  This is also the simplest
analytically and somewhat intuitive in its universality.\footnote{ 
Of course, one could consider other tail or  other crater 
potentials, like a different inverse power of $x$ for the tail 
and/or the $[\cosh^{-2} x]$ potential \cite{vcosh} for the crater.} 

Therefore, we consider  
\bea
V(x) &=&   V_i(x) = \omega^2x^2 -\omega,  ~~~~~~~~~~~~ |x|< x_0 ,    \\
     &=&   V_e(x) = \frac{c(c+1)}{x^2}, ~~~~~~ ~~~~~~~|x|>x_0.
         \label{vsimp}
\eea 
For the external region, which goes to infinity, we take the regular
solution.  From Eqs. (\ref{psiasw}), (\ref{vasw}), and (\ref{vsimp}), 
one has 
\bea
W_e(x)     &=& c\ln x -b , ~~~~~~~~~~~~~ |x|> x_0,  \\
\psi_{0,e}(x)&=& \frac{N~ e^b}{x^c}, ~~~~~~~~~~~~ |x|> x_0.   \label{psiex}
\eea
The constant $b$ is a relative normalization which will affect the location
of the matching point and the relative couplings in the two regions.  We
only need one such constant, so none will be  given for the internal 
region.\footnote{
We are ignoring  the fact that
one needs to handle the sign of $x$ in the negative-$x$ region carefully.  
However, since this problem is symmetric, there is no real concern.  We
just have to note that the phase in the wave function disappears if we let
$(-x\rightarrow x)$ in the original differential equation.}
For the internal region, we {\it assume} for now that one can similarly take
the regular solution.  Therefore,  
\bea
W_i(x) &=& \lfrac{1}{2}\omega x^2,  
              ~~~~~~~~~~~~~~~~~~~~~~~~~~~ |x|< x_0,  \\
\psi_{0,i}(x) &=& N~\exp\left[-\hlf \omega x^2\right],  ~~~~~~~~ |x|< x_0.
       \label{psiin} 
\eea

The superpotential, $W$, and its first
derivative, $W'$, should also be continuous.   
>From Eq. (\ref{psiasw}) and its derivative, 
this requirement is equivalent to the usual 
quantum-mechanical boundary conditions that  $\psi_0(x)$ and 
$\psi_0'(x)$  are continuous.  
Because we want $W$ and $W'$ to be continuous, we must carefully adjust 
the model at the meeting points of the regions, $\pm x_0$.

Performing the matching, one finds that $x_0$ is determined by two 
relationships among $c,~b,~ \omega,$ and $x_0$.  
\bea
 x_0 &=& \exp\left[\frac{1}{2}+\frac{b}{\omega x_0^2}\right]   
      =  \exp\left[\frac{1}{2}+\frac{b}{c}\right], \label{xmatch}    \\
\omega &=& \frac{c}{x_0^2} 
         = \frac{c}{\exp\left[1 +\frac{2b}{c}\right]}.  \label{omatch}
\eea
Therefore, the general result is given by the wave functions of Eqs. 
(\ref{psiex}) and (\ref{psiin}), the matching conditions 
(\ref{xmatch}) and (\ref{omatch}), and the normalization constant
\be
 N(b,c) = \left[\frac{2}{x_0^{(2c-1)}} \frac{e^{2b}}{(2c-1)}
          ~+~x_0~e~\sqrt{(\pi/c)}~
         \mathrm{erf}(\sqrt c)\right]^{-1/2}.  \label{N}
\ee

As indicated, $V(x_0)$ is discontinuous. 
In particular,  the potential starts at 
$V_i(0)= -\omega$ and rises in a parabola until $V_i(x_0) 
= c(c-1)/x_0^2= (c-1)\omega$.  It then 
discontinuously jumps to $V_e(x_0)=c(c+1)/x_0^2= (c+1)\omega$, 
and finally falls off quadratically.  
$V$ being discontinuous means $W''$ is discontinuous which means 
the second derivative of $\psi_0(x_0)$ is discontinuous.  However, 
this is all normal in quantum mechanics.  Recall, for example,  
the particle in a box.  


\section{Examples}

The natural meeting
point of the two functional forms is  given by  
\bea
V_i(x_{0,nat}) &=& V_e(x_{0,nat}), \\
x_{0,nat} &=& \left[\frac{2c(c+1)}{\omega^2}\right]^{1/4}. 
          \label{xnice}
\eea
So why can this not be used, allowing our potential to be continuous?

The reason is our {\it assumption} that the interior solutions are
regular (and therefore, simple, as we want).   
To make the potentials match at the  $x_0$ of Eq. (\ref{xnice}), 
one also would have had to use   
the irregular, interior solution of the Schr\"odinger equation.  
That is the  parabolic cylinder function 
$D_{-1}(i\sqrt{2\omega}~x)$ \cite{lebedev}.  
This would have made $W$ and $\psi_0$ more
complicated, defeating part of the purpose of this exercise.  
With our construction  only the regular Gaussian solution,  
$D_0(\sqrt{2\omega}~x)$, is involved. 



\begin{figure}[p]
 \begin{center}
\noindent    
\psfig{figure=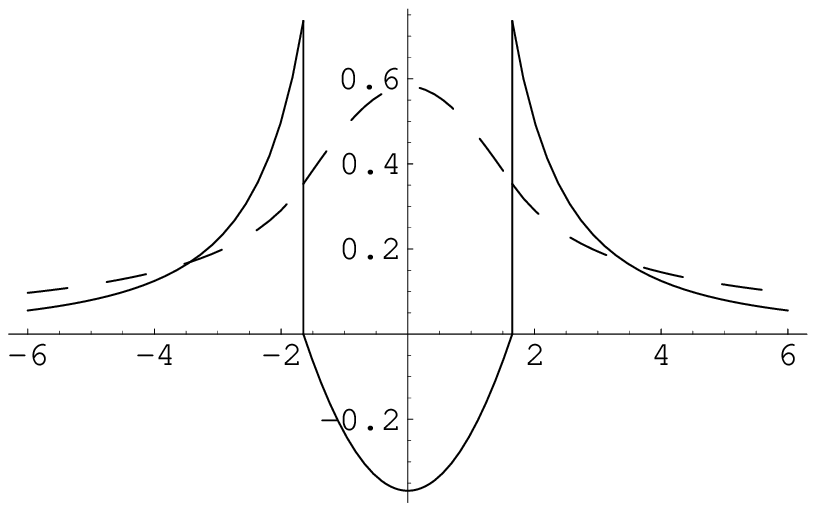,width=4.5in,height=3in}
\caption{For the case $(b,c)=(0,1)$, we show the volcano potential (continuous
  line) and the wave function (dashed line) for the $E=0$ bound state. 
\label{fig:Vulcan1}}

\end{center}
\end{figure} 



\begin{figure}[p]
 \begin{center}
\noindent    
\psfig{figure=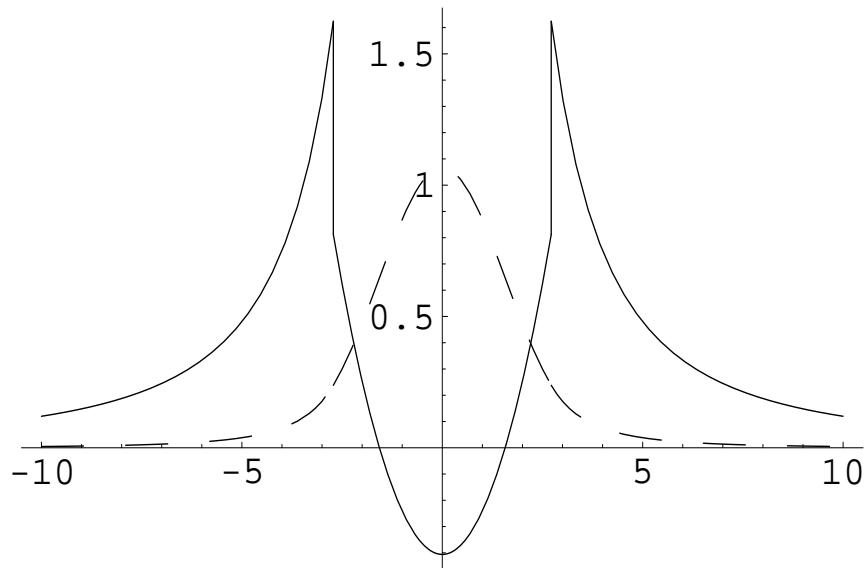,width=4.5in,height=3in}
\caption{For the case $(b,c)=(3/2,3)$, we show the volcano potential 
  (continuous line) and the wave function (dashed line) for the $E=0$
  bound state. 
\label{fig:Vulcan2}}

\end{center}
\end{figure} 


{\bf The case $\mathbf{(b,c)=(0,1)}$:}
Having understood this, for a first  example we look at the most 
elegant and simple special case. It is  
\be
(b,c) = (0,1),  ~~~~~ \Rightarrow ~~~~~ x_0 = e^{1/2}, ~~~\omega = e^{-1},
         ~~~~~~N(0,1)=0.582167.
\ee
In Figure \ref{fig:Vulcan1} we show the potential and the bound, $E=0$, 
ground-state wave function for this $(b,c)=(0,1)$ case.

{\bf The case $\mathbf{(b,c)=(3/2,3)}$:}
For comparison, in Figure \ref{fig:Vulcan2} we show the case 
\be
(b,c) = (3/2,3),  ~~~~~ \Rightarrow ~~~~~ x_0 = e, ~
         ~~~\omega = \frac{3}{e^2},
         ~~~~~~N(0,1)=0.598038.
\ee
Observe that $b\ne 0$ means that $x_0 \ne e^{1/2}$. 

Given the height of the volcano peaks 
(this figure has a different vertical scale than the first), 
one can also ask about resonant
states.  They will be near the energies of the ordinary harmonic oscillator  
excited states, $2n\omega$.  The resonant states  will not be exactly 
at  $2n\omega$ because the 
tail potential evolves  wave functions differently than the full harmonic 
oscillator potential would.  Further, as they
first appear the widths of the resonances are broad. 
They  only become narrow as the volcano wall becomes higher.   
One can approximately say that the number of resonant states is  
\be
n\sim [(c+1)/2], ~~~~~~~~~~~~c\ge 1.
\ee

The case $(b,c)=(0,1)$ has the first resonant state about to appear 
and the case $(b,c) = (3/2,3)$ 
has the second resonant state about to appear.  
These resonant states and the rest of the continuum states 
are solvable, although complicated.  
In the interior one will  obtain parabolic cylinder functions and 
in the exterior one will obtain  Bessel functions.  
In this scenario the resonances do not occur at zero energy. 
(That would have been interesting 
as a model of higher-dimensional gravity whose effects  reappear 
at large distances \cite{reson}-\cite{dvali}.)  
To yield a zero-energy resonance
would necessitate that the potential strictly go to zero at large
distances.  

By contrast, this type of system can yield unbound, zero-energy ground
states.\footnote{  
Unbound $E=0$ solutions were also discussed in Ref. \cite{daboulq}.} 
 For $c<1/2$, the normalization of the wave function 
becomes power-wise divergent.
In particular, note that the zero-energy bound/unbound 
regimes are bordered by the case $c=1/2$.



\begin{figure}[ht]
 \begin{center}
\noindent    
\psfig{figure=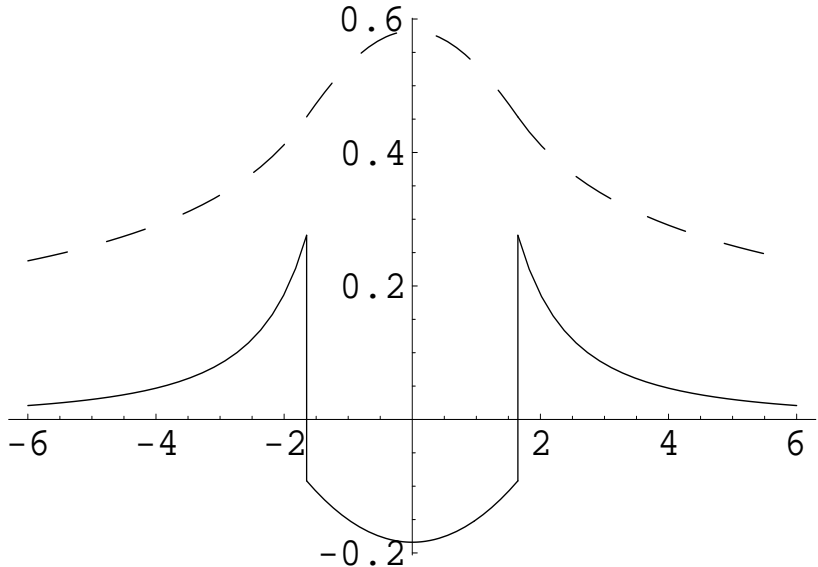,width=4.5in,height=3in}
\caption{For the case $(b,c)=(0,1/2)$, we show the volcano potential 
  (continuous line) and the wave function (dashed line) for the $E=0$
  state, which is logarithmically divergent.  It is arbitrarily set so 
  that $\psi_{0,i}(0) = N(0,1)= 0.582167$.
\label{fig:Vulcan3}}

\end{center}
\end{figure} 


{\bf The case $\mathbf{(b,c)=(0,1/2)}$:}  Here the normalization of the
exterior wave function is  logarithmically divergent. 
That $c=1/2$ is the boundary 
between normalized and unnormalized ground states can also be seen by 
the factor \noindent $2/(2c-1)$ in the normalization constant of 
Eq. (\ref{N}).  To visualize the broad nature of the wave function, 
in Fig. \ref{fig:Vulcan3} we show the case  $(b,c) = (0,1/2)$. 
Since the wave function is not normalized, we set 
$\psi_{0,i}(0) = N(0,1)= 0.582167$. Once again, $x_0=e^{1/2}$.  This is 
an interesting contrast to the first case.
 
Perhaps this $c=1/2$ case is of interest on its own for large scale gravity.  


\section*{Acknowledgements} 

I thank Csaba Cs\'aki, Joshua Erlich, and Tim
Hollowood.   They first asked me about the analytic properties of volcano
potentials and then continuously stimulated the investigation. 
This work was supported by the US DOE.    



\end{document}